\documentclass[journal]{IEEEtran}
\usepackage[utf8]{inputenc}
\usepackage{amssymb}
\usepackage{amsmath}
\usepackage{mathtools}
\usepackage{array}
\usepackage{longtable}
\usepackage{indentfirst}
\usepackage{epsfig}
\usepackage{color}
\usepackage{float}
\usepackage{graphicx}
\usepackage{setspace}
\usepackage{latexsym}
\usepackage{tabu}
\usepackage{multicol}
\usepackage[utf8]{inputenc}
\usepackage[T1]{fontenc}
\usepackage{url}

\newcommand{\bea}{\begin{eqnarray}}
\newcommand{\eea}{\end{eqnarray}}

\begin{document}

\title{Optimum Testing Time of Software using Size-Biased Concepts}
\date{}
\author{Ashis Kumar Chakraborty$^1$, {Parna Chatterjee}$^2$, {Poulami Chakraborty}$^3$\footnote{corresponding author.}, and {Aleena Chanda}$^4$\\
1,2,3 and 4. Statistical Quality Control and Operations Research Division\\
Indian Statistical Institute Kolkata\\
Kolkata-700108, India\\
1. akchakraborty123@rediffmail.com\\
2. parnachatterjee94@gmail.com\\
3. chakraborty.poulami94@gmail.com
\and
4. aleenachanda03@gmail.com\\
}

\maketitle
\begin{abstract}
Optimum software release time problem has been an interesting area of research for several decades now. We introduce here a new concept of size-biased modelling to solve for the optimum software release time. Bayesian approach is used to solve the problem. We also discuss about the applicability of the model for a specific data set, though we believe that the model is applicable to all kind of software reliability data collected in a discrete framework. It has applications in other fields like oil exploration also. Finally, we compare favourably our model with another similar model published recently. We also provide in this article some future possibilities of research work. 
\end{abstract}
\section*{Acronyms}
NB denotes Negative Binomial Distribution.\\
Bin denotes Binomial Distribution.\\
DU denotes Discrete Uniform Distribution.\\ 
MN denotes the Multinomial Distribution.
\section*{Notations}
$N_j$= Cumulative number of runs of the software up to the $j^{th}$ phase , j=1,2, $\cdots$,m. Note that, $N_1$$<$$N_2$$<$$\cdots$\\i.e. $N_j$'s are not independent.\\
$S_{ij}$ = Eventual size of the $i^{th}$ bug identified in the $j^{th}$ phase , i = 1,2,$\cdots$, $n_j$, where $n_j$ is the number of distinct bugs identified upto the $j^{th}$ phase,  j = 1,2,$\cdots$,m \\
$s_{ij}$ = observed size of the $i^{th}$ bug identified in the $j^{th}$ phase \\
$\therefore F_j=\sum_{i=1}^{n_j}S_{ij}; j=1,2,\cdots,m$ i.e. $F_j$ is the total eventual size of the $n_j$ distinct bugs identified up to the $j^{th}$ phase.\\$S_{ij}^{(t)}$ : Eventual size of the $i^{th}$ bug at the $j^{th}$ phase in the time point t\\
$S'_{ij}$ : Eventual size of the $i^{th}$ bug at the $j^{th}$ phase that are either accepted or rejected in the Metropolis Hasttings Algorithm\\
$A_j:(S_{1},S_{2},S_{3},\cdots,S_{j}),S_{0}=0=A_{0}$\\ 
$S_j$ : Total number of bugs detected in the system during the $j^{th}$ phase = $\sum_{i=1}^{n_k}S_{ij}$\\ 
$S_{ij}$ : Eventual size of the $i^{th}$ bug identified in the $j^{th}$ phase. \\ $q_{ij}$ : Probability of detecting the $i^{th}$ bug in the $j^{th}$ phase. \\ $q_{0j}$ : Probability that a bug is not being detected in the $j^{th}$ phase.

\section{Introduction} Software is to be tested before release in the market so that it carries minimum number of errors while using it. It is known that software testing followed by debugging generally improves the reliability of the software. But, a pertinent question raised is, "When to stop testing software?". Several authors ( Dalal and Mallows(1988)\cite{Dalal} , Singpurwalla(1991)\cite{Nozer}, Chakraborty and Arthanari(1994)\cite{CA}, Chakraborty et. al(2019) \cite {doi}, Das et. al(2017)\cite{Sudipta}, Vasanthi et. al(2013)\cite{comp},and many others) have explored different ways of optimizing the time for software release. \\ 
On the other hand, hundreds of models have been proposed to estimate software reliability at a given point of time based on several different assumptions and circumstances. These, in fact, complicated the decision making process of a software developer for appropriately choosing a software reliability model for use. Dalal (2003) provided a selective survey of such models. It was noticed that, for large and critical software, debugging takes place in certain intervals which necessitated development of several different models (Dewanji et. al (2011) \cite{flight control}, Chakraborty et. al (2019) \cite{doi}). For such kind of softwares Nayak(1988) \cite{tk}, Chakraborty(1996) \cite{1996} and a few other authors suggested that software testing data should be collected in a different way in order to get a good estimate of its reliability. Dewanji et. al (2011) \cite{flight control} explained it as a discrete type of data where the data is logged for each input using which the software is being tested and the output is logged as a binary variable indicating whether the software run results in failure or not. Chakraborty and Arthanari (1994) \cite{CA} and Chakraborty et al (2019) \cite{doi} calls it a success when the software fails to provide the right output for an input, because the whole idea of testing the software is to find out as many bugs as possible before it is released in the market. It is known that given an input data, the path that it would take while running the software with the input , is fixed. Whenever, the same input is used, it would traverse the same path and only if there is a bug on that particular path, the given input will be  able to detect that bug. If the bug is not on that path, it is not possible to detect it using that particular input. Since testing is carried out with only a finite and relatively smaller number of inputs, it is difficult to imagine how many inputs would go through a particular bug eventually, if the bug is not fixed. The total input space for which the software would be used will never be known exactly, however, if we can get a good estimate of the cardinality of the input space, it may help us to provide a reliability estimate closer to reality.\par
In this article we introduce a totally new concept in software reliability field and use size-biased sampling concept of Patil and Rao (1978) \cite{RP} to provide a good estimate of potential threat to software failure. We use the newly developed methodology on a testing data collected from a commercial software developer.\par 
The article is organized as follows: In section II, we introduce the new concept of the size of a bug for the first time and develop a Bayesian methodology to get an estimate of the original bug sizes. In Section III we briefly discuss the works of several authors who have explored different ways of optimizing the time for software release. Sections IV mainly deals with the model assumptions used in this model. Section V deals with how to find out optimal time for software testing under this new setup. In section VI we discuss about a software testing data of a commercial software developer with the application of the newly developed size-biased concepts discussed in Section V for the data set. We also compare favourably our model with a similar kind of model developed by Vasanthi and Arulmozhi(2013)(\cite{comp}) in the same section. Further we conclude in Section VII with future directions for research. 

\section{Size-biased concept for Software Reliability }
The size of a bug is defined as the number of inputs that would have eventually passed through the bug, if the bug were not fixed (Chakraborty(1996) \cite{IEEE16}). It is quite natural that a path (in the software) branches into several sub-paths at a later stage. For all these sub-paths, a part of the path is common in the beginning. Imagine that a bug is sitting on the common path and another bug is sitting on one of the several sub-paths associated with the common path. It is quite obvious that the size of the bug, present in the common path is much higher compared to that of the bug in the sub-paths, since all inputs collectively going through each of the sub-paths must be traversing through the common path before entering into a sub-path. The size of a bug also thus may give an indication of how quickly a bug could be identified. If a bigger bug is not detected it would create a potential threat to the functioning of the software, even if there is only one bug. It is simple to understand that the probability of detection of a bug (and hence its fixation) depends on the size of the bug. \par Larger the size of the bug, larger will be the chance of detecting that bug earlier in the testing phase as has been indicated in Chakraborty and Arthanari(1994) \cite{CA}. In fact, Chakraborty and Arthanari(1994) \cite{CA} also have shown that similar concepts are applicable in discovering fields with rich hydrocarbon contents in the field of producing oil and natural gas. It is also clear that a bug which exists in a path that will hardly be traversed by any input, will remain harmless as far as the running of the software is concerned. This brings us to the conclusion that reliability of the software does not depend on just the number of bugs remaining in the software, rather it depends on the positioning of the bugs, particularly the paths on which it exist and whether that path is frequently traversed by inputs which are random in nature as per the user (Littlewood B(1979) \cite{Little}). Hence in order to have a better model for software reliability, our attention would be to find out the total size of the bugs that will remain and not just the number of remaining bugs.\par 
In a discrete software testing framework, when an input is being tested it results in either a failure or a success (finding an error). Testing of software is carried out into many phases where in each phase a series of inputs are tested and results of each testing are recorded as either a success or a failure. After identifying the bugs at the end of testing within a phase, they are debugged at the end of the phase. This process of debugging is known as periodic debugging or interval debugging[Das et al(2016) \cite{IEEE16}]. \par
For testing software, we need to keep in mind certain factors like when we should stop testing, or what will be the criteria of stopping testing etc. If after sufficient testing and debugging most of the bugs remain in the software, then it may result in improper functioning of the software after release in the market. Therefore, a decision to optimize software testing and debugging time is an important part of the development process of software. Even if the number of remaining bugs is smaller, but the total size of the remaining bugs is big, then also the software may fail frequently.

The input space, consisting of all possible inputs to the software can be broadly divided into two subsets of which one consists of all the inputs which will result in a failure (we call it as a success set) and the other subset namely, failure set consists of all the inputs which give expected output. Testing and debugging of bugs consist of several phases in most situations (Dewanji et al) \cite{flight control}. In real life situations, it is quite difficult to debug every time a bug is found as has been assumed by most software reliability models\cite{doi}. It may happen that two inputs have a common path at the beginning due to the presence of some common factors and then each of the inputs branches off to complete the job.

Assume that, as in Dewanji et al. \cite{flight control}, the process of testing flags off as soon as a bug is found and the process is stopped culminating in recording or logging in an incidence of a success. It is easy to understand that the next bug in the path can be detected only after debugging the bug which is detected earlier. Therefore we can assume that the size of a bug which is present at the beginning of a path is much larger compared to the size of a bug present at the end of a sub-path or compared to the bugs present in a path that are hardly traversed by  any input.

\section{Literature review}
Dalal and Mallows(\cite{Dalal})(1988) posed the problem of stopping time for software testing as a trade-off between the cost of testing and the expected loss incurred in case some faults remain in the software after testing is completed. Chao and Mark(\cite{biometrika}) (1993) used a penalty function for the bugs that may remain after the release of the software and modified the cost function to be optimized. Nayak(\cite{tk})(1988), however, suggested a totally different type of data to be collected in the software field in order to get statistically well interpretable software reliability estimates. His paper also is a prelude to the future software testing data that are in the discrete framework. Chakraborty and Arthanari(\cite{CA})(1994) used the concept of discrete framework of software testing to decide optimum time for software testing. Dewanji et. al (\cite{flight control})(2011) considered the ideas provided by Nayak(\cite{tk})(1988), Chakraborty and Arthanari(\cite{CA})(1994) and Chakraborty(\cite{1996})(1996) and developed a model for stopping the testing time of a critical software in order to achieve a stringent reliability target. Chakraborty(\cite{1996})(1996) also propounded a possible approach using the size of the bugs for which data are to be collected using Nayak's(\cite{tk}) (1988) concepts. Zachariah(\cite{optimal stopiing time})(2015) further used the size of the bugs concept of Chakraborty(\cite{1996}) (1996) and defined failure size, at time 't' as a proportion of the cardinality of the input space for which bugs can be identified and a function of the cardinality of the total input space. The optimization with respect to time for software testing is however carried out on a continuous scale by taking a suitable cost function subject to a constraint which is also made continuous in terms of the 'failure size' at time t. \par It was noted by Chakraborty and Arthanari (\cite{CA})(1994), Dewanji et. al(\cite{flight control})(2011) and many others that for critical software the testing data generated are discrete type and unlike most other software reliability models, the debugging takes place in intervals and not immediately after detecting a fault. This is true for large commercial software as well. Das et.al(\cite{Sudipta})(2017) considered several situations under interval debugging and found optimal testing time for software, though their cost function does not include testing cost exclusively. Dalal and Mallows(2008)(\cite{tech}) considered exact confidence on the remaining number of bugs when software testing is completed. Chakraborty et. al(\cite{doi}) (2019) extended the model of Chakraborty and Arthanari(\cite{CA}) (1994)by dropping a very important assumption of immediate debugging after detecting a fault. The idea behind the dropping of the assumption is that most of the large commercial and critical software are debugged after some known or unknown intervals. Vasanthi and Arulmozhi(\cite{comp}) (2013) considered a Bayesian approach to determine the phase till which debugging is needed, such that the probability of detecting any faults after that review is close to 0. They considered Bayesian probability theory to analyze the software reliability model with multiple types of faults. The probability that all faults are detected and corrected after a series of independent software tests and correction cycles is presented. The authors tried to find out the phase, where no class of fault remains in the software system. In this model, the different types of errors have been  grouped as class(1), class(2),$\cdots$,class(k). In the data described in Section 7, we have classified the errors into three classes as: Simple, Complex, Medium. The faults detected in the present review or phase are corrected before starting the next phase. This model has been defined as a discrete time Markov chain model, where a Bayesian approach has been incorporated. The Bayesian approach in this model treats population model parameters as random variables following some distributions. Also, previous information is used to construct a prior model for these parameters. Vasanthi and Arulmozhi considered the prior distribution for the $j^{th}$ phase as $P_{j-1}(v $\textbar$ A_{j-1})$, where, $P_{j}(v $\textbar$ A_{j})$ = conditional probability that after j phases $v$ faults remain undetected given that the data $A_j$ has been observed.\\The likelihood that $S_j$ bugs are detected in the $j^{th}$ phase is taken to be $MN(S_{1j},S_{2j},S_{3j},\cdots,S_{n_jj};S_{j};q_{1j},q_{2j},q_{3j},\cdots,q_{m j},q_{0j})$.\\\\
The posterior probability of $v$ given $A_j$ is : \\$P_{j}(v $|$ A_{j})$ = $\frac{P_{j-1}(S_j+v|A_{j-1})S}{ \sum_{r=0}^{\infty} P_{j-1}(r+S_{j}| A_{j-1})S'}$ \\
where,\\
S=$MN(S_{1j},S_{2j},S_{3j},...\\ \:\:...,S_{n_jj};S_{j};q_{1j},q_{2j},q_{3j},...,q_{m j},q_{0j})$\\
$S'$=$MN(S_{1j},S_{2j},S_{3j},\cdots\\ ...,S_{n_jj},r;{S_{j}+r};q_{1j},q_{2j},q_{3j},...,q_{k+1 j},q_{0j})$\\
$P_{0}(v| A_{0})$ = P(m).  \\
The authors have taken the prior probability that after j phases, v faults remain undetected as Binomial Distribution.\\ 
By inductive hypothesis, it follows that:  \\
$P_{j-1} (v \, | \, A_{j-1}) = {n-S_{1}-... S_{j-1}\choose v}p_{j-1}^v q_{j-1}^{n-S_{1}-... S_{j-1}-v}$, where,\\
$p_{j-1}=\frac{p_{j-2}\times q_{0\, j-1}}{1-p_{j-2} \sum_{i=1}^{n_k} q_{i\, j-1}}$,\\
$q_{j-1}$=$\frac{q_{j-2}}{1-p_{j-2}\sum_{i=1}^{n_k}q_{i j-1}}$,\\ n=$\sum_{j=1}^{m}n_j$ \\
and $p_1$+$q_1$=1.\\
Therefore, for the $j^{th}$ review, assuming the prior probability as $P_{j-1}(v $|$ A_{j-1})$: \\ $P_j(v|A_j)$ = ${n-S_{1}-S_{2}-... S_{j}\choose v}$ \Bigg($\frac{p_{j-1} q_{0 j}}{1-p_{j-1}\sum_{i=1}^{n_k}q_{i j}}\Bigg)^{v}$  $\times \Bigg({\frac{q_{j-1}}{1-p_{j-1}\sum_{i=1}^{n_k}q_{i j}}}\Bigg) ^{n-S_{1}-S_{2}-... S_{j}-v}$ \\ 
with, $p_{j}$ = $\frac{p_{j-1}q_{0 j}}{1-p_{j-1}\sum_{i=1}^{n_k}q_{i j}}$,\\ $q_{j}$=$\frac{q_{j-1}}{1-p_{j-1}\sum_{i=1}^{n_k}q_{i j}}$.\\ $P_j(v|A_j)$ = ${n-S_{1}-S_{2}-... S_{j}\choose v}p_{j}^v q_{j}^{n-S_{1}-S_{2}-... S_{j}-v}$. \\
The problem is to find the $j^{th}$ phase for which $P_j(v| A_j)\rightarrow 1,\forall$ j=1,2,...,m. All the aforementioned works as well as other existing works have not taken into consideration the fact that the probability of detection of a bug depends on the size of the bug. To address this issue we introduce the concept of size- biased sampling in our paper. Despite all these attempts, the optimum time for testing software remains an interesting problem to the researchers. In this article we provide a completely new way of modeling the phenomena by taking recourse to size-biased sampling indicated by Chakraborty (\cite{1996})(1996).
 \section{Model Assumptions}It is assumed that whenever an input finds out a bug in the first instance,  it signals a success indicating that a bug is found. The testing group will then possibly try out with a different input. It is to be noted that the software industry spends about 50-60\% of its development cost for testing alone. The aim of the testing group is to find out as many bugs as possible during testing itself, so that the released software will have high reliability.

In a path when a bug is identified by an input for the first time it can be defined as level 1 bug. It is reasonable to assume that in a path level 1 bugs will be identified first, if that path is traversed by an input selected for testing. After debugging in each phase the level 1 bugs identified during that phase would be debugged. In a subsequent phase, if another input is chosen which traverses the same path and identifies another bug in the path, then that bug is called as level 2 bug and so on. Our objective is to find out optimally when to stop testing the software using interval debugging and the concepts of size of a bug.\\
We assume that, if we continue testing till all the bugs are identified and debugged, we need m phases of testing where in each phase $N_i$, i=1,2,$\cdots$,m inputs will be used. We also assume that all the bugs that are identified in a particular phase are debugged before the start of a new phase. Let us suppose that, the software company has testing data up to the $k^{th}$ phase. This means the bugs are detected and debugged only up to the k phases. In general, it is expected that the number of distinct bugs identified in the $k^{th}$ phase shall be greater than that in the $(k+1)^{th}$ phase. Further, it is also expected that, the total size of the bugs should decrease as debugging is carried through the phases provided that the group of testers is efficient. The problem now boils down to, given some $\epsilon$, however small, stop testing after l phases, where l satisfies the condition that the total eventual size of the bugs after l phases of testing and debugging is less than $\epsilon$. This, in a way would ensure that after $l^{th}$ phase of testing, the software has achieved certain reliability as prescribed by the customer. 
\section{The Model}
Our main objective is to estimate the total eventual size of the remaining bugs after each phase in order to determine the optimal time for testing. If the predicted value of the total eventual size of the bugs in the $j+1^{th}$ phase becomes less than $\epsilon$, fixed earlier, then we should stop at the $j^{th}$ phase itself.\\ Now estimating the remaining size of bugs is a difficult proposition. It is next to impossible to get exact value of the total eventual size of the bugs up to the final (eventual) phase. So, we will try to find an estimate of the total eventual size of the bugs for each phase and stop at that phase when the total predicted eventual size of the bugs for the very next phase is very small.\\ 
Let us assume that $N_j|N_1,N_2,\cdots,N_{j-1},F_1,F_2,\cdots,F_{j-1}\sim g_{N_j| N_1,N_2,\cdots,N_{j-1},F_1,F_2,\cdots,F_{j-1}}(N_j)$.\\
It is to be noted, as discussed in Section II, that the bugs situated at the beginning of the path are expected to have higher probability of detection compared to the bugs that are towards the end of a path. Hence, the mass function h will be modified according to the size of the bug.\\ 
${h_S}_{ij}(s_{ij})=\frac{f_{S_{ij}}(s_{ij})\times s_{ij}}{E_{f}(S_{ij})}$ , where ${f_S}_{ij}(s_{ij})$ is the original distribution of the size of bug.\\ 
The likelihood function of $S_{ij}$ is given by \\
$L(S_{1j}, S_{2j},\cdots,S_{nj{_j}}|N_1,N_2,N_3,\cdots,N_m)$\\
$=\prod_{j=1}^{m}g_{N_j}|N_1,N_2,\cdots,N_{j-1},F_1,F_2,\cdots,F_{j-1}(N_j)$\\ 
The posterior distribution of $S_{ij}$; i=1,2,..,$n_j$;j=1,2,...,m is given by: 
$\pi_{S_{1j},S_{2j},...,S_{nj_{j}}}(S_{1j},S_{2j},...,S_{nj_j}|   N_1,N_2,...,N_m)\\ 
\text{ which is proportional to}\\
L(S_{1j},S_{2j}, .....,S_{nj_j}|N_1,N_2, N_3, .., N_m)\prod_{i=1}^{n_j}{h_S}_{ij}(s_{ij})$.\\
If the posterior distribution has a closed form, then the estimate for total eventual size of the bugs in a phase can be obtained from the posterior distributions using Gibbs Sampling, But if the posterior distribution does not have a closed form, then Metropolis Hastings algorithm has to be used in order to obtain the estimates. We now consider various distributional choices for $N_j$ and $S_{ij}$. 
\subsection{Distributional Choices}$N_j$ is the cumulative number of runs that would lead to $F_j$, i.e. the total eventual size of $n_j$ distinct bugs identified up to the $j^{th}$ phase.\\
Let $N_1\;\thicksim\;NB(F_1\:,\:p_1)$\\
$N_2 | N_1,F_1 \thicksim NB(F_2-F_1,p_2)$\\ 
In general, $N_k|N_1,N_2,\cdots,N_{k-1},F_1,F_2,\cdots,F_{k-1}\;\thicksim\;NB(F_k-\sum_{i=1}^{k-1}F_i,p_k)$\\ 
and $N_m|N_1,N_2,....,N_{m-1},F_1,F_2,\cdots,F_{m-1}\;\thicksim\;NB(F_m-\sum_{i=1}^{m-1}F_i,p_m)$\\

The likelihood function $(S_{1j},S_{2j},\cdots,S_{n_jj},p_1,p_2,\cdots,p_m)$ is of given by:
\begin{eqnarray}{\label{one}}
 L(S_{1j},S_{2j},..,S_{n_jj},p_1,p_2,p_3,..,p_m|N_1,N_2, N_3, ..,N_m) \nonumber \\
={N_1+F_1-1\choose N_1} p_1^{N_1} (1-p_1)^{F_1} \times \nonumber\\
{N_2+F_2-F_1-1\choose N_2} p_2^{N_2} (1-p_2)^{F_2} \times \nonumber\\ {N_3+F_3-F_2-F_1-1\choose N_3} p_3^{N_3} (1-p_3)^{F_3} \times ...\times \nonumber\\ 
{N_m+F_m-\sum_{i=1}^{m-1}F_i-1\choose N_m}p_m^{N_m}(1-p_m)^{F_m-\sum_{i=1}^{m-1}F_i} \nonumber\\
\end{eqnarray}
Our objective is to find out that phase for which the total size of the bugs is less than a very small given quantity $\epsilon$. In other words, our objective is to determine k-1 for which $\sum_{i=1}^{n_k}S_{ik} < \varepsilon$, i= 1,2,$\cdots$ where $\varepsilon>0$ is a very small quantity.
Since the likelihood function is of complex form  estimating $S_{ij}$ becomes difficult in general. Hence, for simplicity we consider three different distributions for the size of the bugs and use Bayesian technique to estimate the total eventual size of the bugs in different phases. These three distributions have been mainly chosen on the basis that size of the bugs always takes non negative integer values. Consideration of the Binomial distribution as the distribution of the size of the bugs is justifiable based on the fact that the appearance of a bug may be termed as a success in a sequence of independent trials and not detecting a bug in a trial maybe considered as a failure(Chakraborty and Arthanari)(1994)(\cite{CA}). Further, since the size of the bugs can be theoretically large and they are non-negative integers, the distributional choices for the size of bugs can also be taken to follow Negative Binomial or Poisson distribution.

 With each one of the target distributions given above, we consider three choices of proposal distributions where the support of each of these distributions is (0, $\infty$) which covers the support of $S_{ij}$. The three choices of proposal distributions are as follows : 1. Geometric Distribution, 2. Negative Binomial Distribution, 3. Poisson Distribution.
 As a result, altogether nine combinations of distributions can be formed. However, in this article we describe the combination where the distributional choice of the size of bugs is taken as Binomial and the proposal distribution is taken as Poisson, since this combination gives highest rate of convergence compared to the others.
\subsection{Choice of Prior}The choice of prior should always be justifiable based on arguments. On the other hand, the prior information is rarely rich enough to define a prior distribution exactly. So, it is necessary to include this uncertainty in the model. We will use Hierarchical Bayes' Analysis in order to improve the robustness of the estimates of the total eventual size of the bugs. \\ $S_{ij}$ $\thicksim$ Bin($n_{ij}$ , $t_{ij}$) ; i = 1,2,$\cdots$, $n_j$ ; j = 1,2,$\cdots$ m. \\
Let us assume,\\
$ t_{ij}\;\thicksim\;Beta(a_{ij},b_{ij}),\: where\: a_{ij}\: and\: b_{ij}\: are\: known $.\\
$ p_j\;\thicksim\;Beta(\alpha_j\:,\:\beta_j)\:,\:j=1,2,\cdots,m. $\\
Now $\alpha_j$ and $\beta_j$ are unknown, so we estimate them as follows.\\
We\;know,\\
\begin{equation}
E\:(p_j|\alpha_j\:,\:\beta_j)\;=\;\frac{\alpha_j}{\alpha_j+\beta_j}\;=\;\mu_j, \: say
\end{equation}
and
\begin{equation}
 Var(p_j|\alpha_j,\:\beta_j)\;=\;\frac{\alpha_j\beta_j}{(\alpha_j+\beta_j)^2(\alpha_j+\beta_j+1)}\;=\;\sigma_j^2
\end{equation}
where,\: $\mu_j$ $\thicksim$U(0 , 1) and $\sigma_j^2$\textbar$\mu_j$ $\thicksim$ U(0 , $\mu_j$(1-$\mu_j$) ).\\
The estimates of $\alpha_{j}$ and $\beta_j$ are obtained by solving Equations (2) and (3), i.e.\\ 
$\widehat{\alpha_j}$ = $\mu_j [\dfrac{\mu_j (1-\mu_j)}{\sigma_j^2}-1]$ and $\widehat{\beta_j}$ = $\dfrac{\widehat\alpha_j}{\mu_j} (1-\mu_j)$.\\
$F_j$ = $\sum_{i=1}^{n}S_{ij}$ ; j = 1,2,$\cdots$, m.\\
Here n is unknown, so we assume,\\n $\thicksim$ DU(1, 2, $\cdots$ , k), where k is fixed.\\ 
The mass function of $n_{ij}$ is defined as ,\\
$\;n_{ij}\:=m_{ij}$, with probability $\frac{m_{ij}}{\sum_{j=1}^{m}m_{ij}}$.

\subsection{Posterior Distributions}The joint posterior distribution under the above setup is\\
\begin{eqnarray}
\pi(A|B)\propto {N_1+F_1-1\choose N_1}{p_1^{N_1}}{(1-p_1)^{1-F_1}}\times \nonumber\\
{N_2+F_2-F_1-1\choose N_2} {p_2^{N_2}} {(1-p_2)^{F_2-F_1}}\times \cdots\times \nonumber\\ 
{N_m+F_m-{\sum_{i=1}^{m-1}F_i}-1\choose N_m} {p_m^{N_m}}(1-p_m)^{F_m-{\sum_{i=1}^{m-1}F_i}}\times \nonumber\\ 
\prod_{i=1}^{m}p_i^{\alpha_{i-1}}(1-p_i)^{\beta_{i-1}}\prod_{i=1}^{n}s_{ij}{{n_{ij}\choose s_{ij}}} t_{ij}^{s_ij} (1-t_{ij})^{n_{ij}-s_{ij}} 
\end{eqnarray}
where\\
A=$S_{1j},S_{2j},\cdots,S_{n_jj},p_1,p_2,..,p_m,t_{1j},t_{2j},\cdots,t_{nj},n_{1j},n_{2j},\cdots,n_{nj}$\\
B=$N_1,N_2,\cdots,N_m,\alpha_1,\alpha_2,\cdots,\alpha_m,\beta_1,\beta_2,\cdots,\beta_m$\\
The posterior of $p_i$ is given by :
\begin{eqnarray}
\pi_{p_{i}}(p_{i}|S_{1j},S_{2j},\cdots,S_{n_jj}N_1,N_2,\cdots,N_m, \alpha_1,\alpha_2,..\nonumber\\
..,\alpha_m,\beta_1,\beta_2,\cdots,\beta_m) \nonumber\\
\propto p_i^{N_i+\widehat{\alpha_i}-1} (1-p_i)^{F_i-\sum_{j=1}^{i-1}F_j+\widehat{\beta_i}-1}
\end{eqnarray}
The posterior of $t_{ij}$ is given by :
\begin{eqnarray}
\pi_{t_{ij}}(t_{ij}|N_1,N_2,...,N_m, a_{ij},b_{ij},s_{ij},n_{ij}) \nonumber\\
\propto t_{ij}^{s_{ij}+a_{ij}-1}(1-t_{ij})^{n_{ij}-s_{ij}+b_{ij}-1} 
\end{eqnarray}

The posterior of $S_{ij}$ is given by :
\begin{eqnarray}
\pi_{S_{1j},S_{2j},\cdots,S_{n_jj}}(C|D ) \propto {N_1+F_1-1\choose N_1}\times \nonumber\\ 
{N_2+F_2-F_1-1\choose N_2}\times \cdots {N_m+F_m-\sum_{i=1}^{m-1}F_i -
1\choose N_m}  \nonumber\\  
\times (1-{t_{ij}})^{n_{ij}-s_{ij}} (1-p_1)^{F_1} (1-p_2)^{F_2-F_1}(1-p_3)^{F_3-F_2}  \nonumber\\ 
\times ...(1-p_m)^{F_m-\sum_{i=1}^{m-1}F_i}\times{\prod_{i=1}^{n} s_{ij} {n_{ij}\choose s_{ij}}{t_{ij}}^{s_{ij}}}
\end{eqnarray}
where C=$S_{1j},S_{2j},\cdots,S_{n_jj}$\\ 
D= $N_1,N_2,\cdots,N_m,\alpha_1,\alpha_2,\cdots,\alpha_m,\beta_1,\beta_2,\cdots$\\
$\cdots,\beta_m,t_{1j},t_{2j},\cdots,t_{nj},n_{1j},n_{2j},\cdots,n_{nj} \cdots$\\

Since, the posterior distribution of $S_{ij}$, i=1,2,$\cdots$,$n_j$; j=1,2,$\cdots$,m does not have a closed from, we apply Metropolis Hastings Algorithm to get estimates of $S_{ij}$.\\
The algorithm is as follows:\\
\\ Step \textcircled{1}\\ Initialize the starting state $S_{ij}^{(t)}$ at t = 0.\\
Step \textcircled{2}\\ Draw a sample $S_{ij}'$ from the proposal q($S_{ij}'|S_{ij}^{(t)}$)\\ 
Step \textcircled{3}\\Decide whether we should accept the new state by computing $\alpha$ which is given by,\\$\alpha=\frac{\pi_{S_{ij}^{'}}(s_{ij}^{'})\:q(s_{ij}^{(t)} | s_{ij}^{'})} {\pi_{S_{ij}^{(t)}}(s_{ij}^{(t)})\:q(s_{ij}^{'} | s_{ij}^{(t)})}$\\
Accept the new state with a probability of:\\
$A(S_{ij}^{'}|S_{ij}^{(t)}) = min(1,\alpha)$, i.e.\\ 
\textcircled{a} Generate a uniform random number u$\epsilon$ (0,1) \\
\textcircled{b} If $u\leqslant A(S_{ij}^{'}|S_{ij}^{(t)}),\: then \: S_{ij}^{(t+1)}= S_{ij}^{'}$\\
\textcircled{c} If $u > A(S_{ij}^{'}|S_{ij}^{(t)}), \: then\:  S_{ij}^{(t+1)}=S_{ij}^{(t)}$\\
Repeat Steps \textcircled{b} and \textcircled{c} until convergence.\\
Since the support of Poisson distribution is $(0,\infty)$ which covers the support of $S_{ij}$, we take Poisson to be our proposal distribution, i.e. $q(S_{ij})=\frac{e^{-\lambda_{ij}}\lambda_{ij}^{s_{ij}}}{s_{ij}!}$\\
Here,
\begin{equation}
  \alpha= \left(\frac{G}{H}\right)   
\end{equation}
where, \\G\:=\:$\lambda^{s_{ij}^{(t)}} s_{ij}^{'}! {N_1+F_1^{'}-1 \choose N_1} (1-p_1)^{F_1^{'}} {N_2+F_2^{'}-F_1^{'}-1 \choose N_2}\times $\\
$(1-p_2)^{F_2^{'}-F_1^{'}}....{N_m+F_m^{'}-\sum_{i=1}^{m-1}F_i^{'}-1 \choose N_m}(1-p_m)^{F_m^{'}-\sum_{i=1}^{m-1}F_i^{'}} \times $\\
$\prod_{i=1}^{n}s_{ij}^{'}{n_{ij} \choose s_{ij}^{'}}\:t_{ij}^{s_{ij}^{'}}(1-t_{ij})^{n_{ij}-s_{ij}^{'}}$\\
H\: =\: $\lambda^{s_{ij}^{'}}\: s_{ij}^{(t)}! {N_1+F_1^{(t)}-1 \choose N_1} (1-p_1)^{F_1^{(t)}} {N_2+F_2^{(t)}-F_1^{(t)}-1 \choose N_2} \times $\\
$(1-p_2)^{F_2^{(t)}-F_1^{(t)}}....{N_m+F_m^{(t)}-\sum_{i=1}^{m-1}F_i^{(t)}-1 \choose N_m} \times $\\
$(1-p_m)^{F_m^{(t)}-\sum_{i=1}^{m-1}F_i^{(t)}}\prod_{i=1}^{n}s_{ij}^{(t)}{n_{ij} \choose s_{ij}^{(t)}}\:t_{ij}^{s_{ij}^{(t)}}(1-t_{ij})^{n_{ij}-s_{ij}^{(t)}}$\\

The estimates of the total eventual size of the bugs in each phase obtained are given in Table II in section VI.\\

Our objective is to determine k-1 for which $\sum_{i=1}^{n_k}S_{ij} < \varepsilon,i= 1,2,....n_k$ where $\varepsilon>0$ is a very small quantity.If information is not available for a particular phase, estimates of the size of the bugs for that phase becomes complicated. We employ here a predictive method called Predictive Kernel Density Estimation Method, in order to obtain the estimates of the size of the bugs at later phases.\\
\subsection{Predictive KDE}
In order to estimate the total size of the bugs for the phase for which no information is available, we consider temporally weighted kernel density models (Porter and Reich(2012))(\cite{kernel}). The temporally weighted kernel is given by : \\ $\widehat{f}(s|t) = \sum_{j=1}^{t}w_j(t).K(||S-S_j||;h)$\\where t is the number of past phases and $w_j(t)$ is the weight of the the j$^{th}$ phase at time t. We take $w_j(t)$ $>$ 0; j=1,2,$\cdots$,m and $\sum_{j=1}^{t}w_i(t)$=1 ,such that $\int \widehat{f}(s|t) ds= 1 $.
\subsubsection{Choice of Kernel}
Since the total size of the bugs cannot be negative and can take any value between (0,$\infty$) we choose K(u;h) as the isotropic 2D Gaussian Kernel, i.e.\\$K_h(u)$ = $\dfrac{1}{2\pi h^2}e^{-\dfrac{u^2}{2 h^2}}$.\\Also, since the exact time for which the j$^{th}$ event, namely $j^{th}$ phase, takes place is not known, we take a time window for the event j to be [$v_j,\eta_j$]. The weight is thus given by:\\$w_j(t) \propto E[g(t-t_j$)]\\ = $\dfrac{1}{\eta_j-v_j}\int_{v_j}^{\eta_j}g(t-u)du$.\\=$\dfrac{1}{\eta_j-v_j}[G(t-v_j)-G(t-\eta_j)]$, where g(t-$t_j$) is a temporal kernel and G(u) is the cumulative distribution function of g(u).Here $v_j$ is the the earliest possible time and $\eta_j$ is the latest possible time for the event j. Thus, $S_j$ falls within the time interval $[v_j,\eta_j]$ ; j=1,2,$\cdots$,m. We choose the exponential distribution as the temporal weighting function due to its simplicity and familiarity and to emphasize that it is a one-sided (predictive) Kernel since the phase numbers are always increasing and cannot be negative. The $j^{th}$ event is predicted based on the estimates of the previous j-1 events. Hence, for a given bandwidth h ,the prospective kernel estimate of the conditional density for the total eventual bug size for the $j^{th}$ phase is estimated on the basis of the following criterion:\\$S_{k-1} > S_k$ where $S_k$ = $\sum_{i=1}^{n_k}S_{ij}$. 
\subsubsection{Bandwidth Selection}
 A distance measure considered between $\hat{f}$ and f and termed as Integrated Squared Error (ISE) is given by,\\
 $d_{1}(h)=\int{(\hat{f_h}-f)^2(x)dx}\\\:\:\:\:=\int{\hat{f_h}^2(x)dx}-2\int{(\hat{f_h}f)(x)dx}+\int{f^2(x)dx}$\\
Note that, $ f^2(x)$ does not depend on h.Further, we observe that $\int{(\hat{f_h}f)(x)dx}=E_X[\hat{f_h}(X)]$, where the expectation is understood to be computed with respect to an additional and independent observation X. For estimation of this term define the leave one out estimate,\\
$E_X[\hat{f_h}(X)]= n^{-1}\sum_{i=1}^{n}\hat{f_{h,i}}(X_i)$, \\ 
where $\hat{f_{h,i}}(X_i)$ is an estimate based on the subset $\{X_{j}\}_{j\neq i}$\\
Using this estimate we determine a bandwidth h minimizing,\\ $CV(h)=\int{\hat{f_h}^2(x)dx}-\frac{2}{n} n^{-1}\sum_{i=1}^{n}\hat{f_{h,i}}(X_i)$

\section{Application}
\subsection{Data}
The data set consists of 8757 test inputs detailed with build number , case id, severity, cycle, result of test, defect id etc. In this data, the severity of a path is broadly divided into three categories namely, simple, medium and complex depending on the effect of the bug if it is not debugged before marketing the software. The data has four cycles namely Cycle 1, Cycle 2, Cycle 3 and Cycle 4, which is equivalent to the different phases of testing we have referred to section III. After each cycle, the bugs that are identified during the cycle are debugged as mentioned in the section III. If during testing of a software no bug is found then the results are stated as “executed successfully”, but if there is a bug identified during testing then the results are divided into several categories viz. functionality, blocked, deferred, fail, no run and not in scope. Functionality means the bug, if not corrected would affect some functions in the software, whereas blocked, deferred etc. are characteristics of a bug which indicates the possible actions to be taken by the software developers in that particular cycle.  If the defect header of two bugs are same it means that those two bugs are generated from the same source with different Id. However, we are interested only with the observed size of the bugs that has gone through a particular defect Id in a particular cycle.
Hence, the original data is summarized in a form of the cycle number , defect header, defect Id along with the number of inputs. A sample of the data thus collected is given in table 1. From these data the estimates of the population size for each identified bug is calculated using the methodology developed in section V.
\begin{table} [H]
\caption{\textbf{A sample from the Data}}
\begin{tabular}{|p{1.5cm}|p{1.5cm}|p{1.5cm}|p{1.5cm}|} 
\hline \textbf{Cycle} & \textbf{Defect Header}& \textbf{Defect id}& \textbf{Size}\\
\hline
1 & 2 & 3 & 1\\\hline 1&5&6&3
\\\hline1 & 5 & 7 & 13\\\hline
2 & 13 & 31 & 2\\\hline
2 & 15 & 31 & 16\\\hline
3 & 14 & 10 & 1\\\hline
3 & 23 & 4 & 8\\\hline
3 & 25 & 2 & 1\\\hline
4 & 5 & 13 & 4\\\hline
4 & 42 & 4 & 2\\\hline 
\end{tabular} 
\end{table}

\subsection{Results}
\begin{table} [H]
\caption{\textbf{ESTIMATES OF THE TOTAL EVENTUAL SIZE OF THE BUGS IN EACH PHASE}}
\begin{tabular}{|p{3cm}|p{4cm}|} 
\hline \textbf{PHASE} & \textbf{TOTAL EVENTUAL SIZE}\\ \hline
1 & 34007\\ \hline
2 & 36157\\\hline
3 & 57738\\\hline
4 & 11409 \\\hline
5 & $6.9\times 10^-{10}$ \\\hline
\end{tabular} 
\end{table}

Using the Predictive Kernel density Estimation technique the total eventual size of the bugs in the fifth phase has been estimated to be $6.9\times 10^-{10}$ which is approximately equal to 0. Hence, according to the data one can consider the fourth phase to be the optimal testing time of software,that is one can stop the process of debugging in the fourth phase itself.\\
It can also be noticed that there is an increase in the total eventual size of the bugs from the first to the second phase as well as from the second to the third phase. This goes against our assumption. Another noticeable thing is that, this increase in size from the second to the third phase is very high . It is not guaranteed that whenever an input passes through the path a bug will always be detected (Success). Hence, though the number of inputs provided in the second phase is more than that in the third phase , it may so happen that the inputs in the third phase may have passed through the paths on which the bugs were present, whereas in the second phase the inputs may have traversed through the paths where bugs were not present. Thus, the estimate of total eventual size of the bugs have increased sharply from the second to the third phase, even though the number of inputs were less.\\
This possibly shows that the testing group is  getting more and more experienced as the number of phase increases as is expected. Further, on comparing the proposed model with the model proposed by Vasanthi and Arulmozhi(2013), using Bayes' Factor it has been found out that our model is approximately 67\% times better.
\section{Conclusion}
In this article, we propose a completely new approach for solving the problem of determining optimal time for software release. The approach of size-biased sampling seems to be more realistic and is based on the concepts of Patil and Rao(1978). Since, the theoretical solution for the estimates of $S_{ij}$ does not have a closed form, we propose to take recourse to some specific distributions of the parameters and solve the problem using Bayes' Method.\par The proposed approach is applied to a software testing data of a large commercial software. Our approach is also compared with a similar approach proposed by Vasanthi and Arulmozhi(2013). It is shown that our proposed model works better 67\% times out of a trial run of 10,000 and the Relative MSE also is comparatively less (0.61 $\%$) for our proposed model. Thus, it can be said that the proposed size-biased model is quite efficient. \par Finding out reliability of the software when it is ready for release under the present set up is an interesting work that we are looking into.


\begin{thebibliography}{9}
\bibitem{doi} A.K. Chakraborty, G.K. Basak, and S. Das, "Bayesian optimum stopping rule for software release", DOI no : 10.1007/s12597-018-00353-0,2019.
\bibitem{1996}  A.K. Chakraborty, "Software Quality Testing and Remedies", PhD thesis, 1996. 
\bibitem{biometrika} A. Chao and  C.K. Mark, "Stopping Rules and Estimation for Recapturing Debugging with Unequal Failure Rates".{\it Biometrika}, vol. 80, no. 1, pp. 193-201, 1993.
\bibitem{CA}A.K. Chakraborty and T.S. Arthanari, "Optimum testing time for software under an exploration model", \it{Opsearch},vol. 31, pp. 202-214, 1994.
\bibitem{tech} S.R. Dalal and C.L. Mallows, "Optimal Stopping With Exact Confidence on Remaining Defects",\it{Technometrics},vol. 50, no. 3, pp. 397-408, 2008. 
\bibitem{Dalal}S.R. Dalal and C.L. Mallows, "When should one stop testing software?",\it { Journal of the American Statistical Association}, vol. 83, no. 403, pp. 872-879, 1988.
\bibitem{SR}S.R. Dalal, "Software reliability models: A selective survey and new directions", in \it{Handbook of Reliability Engineering}, ed. H. Pham,  London: Springer, pp. 201-2011, 2003.
\bibitem{Sudipta} S. Das, D. Sengupta, and A. Dewanji, "Optimum release time of a software under periodic debugging schedule", \it{ Communications in Statistics-Simulation and Computation}, pp. 1-19, 2017.
\bibitem{IEEE16}S. Das, A. Dewanji, and A.K. Chakraborty, "Software Reliability Modeling with Periodic Debugging Shedule". {\it IEEE Transactions On Reliability}, vol. 65, no. 3, pp. 1449-1456, 2016.
\bibitem{flight control}A. Dewanji, D. Sengupta, and A.K. Chakraborty, " A discrete time model for software reliability with application to a flight control software", \it{Applied Stochastic Models in Business and Industry}, vol. 27, pp. 723-731, 2011.
\bibitem{Little}B. Littlewood,"Software Reliability Model for modular program structure", \it{IEEE Transactions on Reliability}, vol. 28, no. 3, pp. 241-246, 1979. 
\bibitem{tk}T.K. Nayak, "Estimation of Population Size by recapture sampling", \it{Biometrika}, vol. 75, no. 1, pp. 113-120, 1988.
\bibitem{RP}G.P. Patil and C.R. Rao,  "Weighted Distributions and Size-Biased Sampling with Applications to Wildlife Populations and Human Families", \it{Biometrics} vol. 34, no. 2, pp. 179-189, 1978.
\bibitem{kernel} M.D. Porter, and B.J. Reich, "Evaluating temporally weighted kernel density methods for predicting the next event location in a series", \it{Annals of GIS},vol. 18, no. 3, pp. 225-240, 2012.
\bibitem{Nozer} N.D. Singpurwalla, "Determining an optimal time interval for testing and debugging software",\it{ IEEE Transactions on Software Engineering},vol. 17, no. 4, pp. 313-319, 1991.
\bibitem{comp}T. Vasanthi and G. Arulmozhi, "Software reliability estimation using Bayesian approach",{\it International Journal of Quality \& Reliability Management}, vol. 30, no. 1,pp. 97-107, 2013.
\bibitem{testisng srategies}B. Zachariah, "Analysis of Software Testing Strategies Through Attained Failure Size",{\it IEEE Transactions On Reliability}, vol. 61, no. 2, pp.569-579, 2012.
\bibitem{optimal stopiing time}B. Zachariah,"Optimal stoping time in software tetsing based on failures size approach",{\it Ann Oper Res}, vol. 235, pp. 771-784, 2007.
\bibitem{failure size proportional model} B. Zachariah and  R.N. Rattihalli,"Failure Size Proportional Models and an Analysis of Failure Detection Abilities of software Testing Strategies,\it{ IEEE Transactions On Reliability}, vol. 56, no. 2, pp.246-253, 2007.
\end {thebibliography}
\end{document}